%%%%%%%%%%%%%%%%%%%%%%% use TeX %%%%%%%%%%%%%%%%%%%%%%%%%%%%%%%%%
\magnification=1200
\overfullrule=0pt
\baselineskip=20pt
\parskip=0pt
\def\dag{\dagger}
\def\del{\partial}

\def\a{\alpha}             
              
\def\g{\gamma}        
\def\d{\delta}        
\def\e{\epsilon}           
\def\z{\zeta}              
\def\j{\eta}

\def\k{\kappa}     
      
\def\m{\mu}	   \def\M{M}        
\def\n{\nu}                
\def\x{\xi}              
                  
\def\p{\pi}        \def\P{\Pi}     
\def\r{\rho}               
\def\s{\sigma}

\def\f{\phi}            
\def\h{\chi}
\def\y{\psi}

\def\Y{{\mit\Psi}}      

\def\w{\omega}     
   
\def\br{\langle}
\def\ke{\rangle}
\def\ve{\vert}
\def\inf{\infty}
\def\Winf{$W_{\infty}\  $}
\def\winf{$w_{\infty}\  $}
\def\to{\rightarrow}

\def\zbar{\bar{z}}
\def\zintm{d^2 z e^{-|z|^2}}
\def\zintmp{d^2 z' e^{-|z'|^2}}
\def\to{\rightarrow}
\def\tr{{\rm tr}}

\def\ba{{\cal A}}
\def\bF{{\cal F}}

\def\vf{\varphi}
%%%%%%%%%%%%%%%%%%%%%%%%%%%%%%%%%%%%%%%%%%%%%%%%%%%%%%
{\settabs 5 \columns
\+&&&&CCNY-HEP-96/4\cr
\+&&&&February 1996\cr}
\bigskip
\centerline{\bf \Winf AND \winf GAUGE THEORIES AND CONTRACTION}
\bigskip\bigskip
\centerline{ An. Kavalov and B. Sakita$^*$}
\bigskip
 
\centerline{ Physics Department, City College 
of the City University of New York}
\centerline{ New York, NY 10031}
\bigskip
\bigskip \bigskip
\centerline{\bf Abstract}

We present a general method of constructing 
\Winf and \winf
gauge theories in terms of $d+2$ dimensional 
local fields.
In this formulation the \Winf gauge theory Lagrangians
involve non-local 
interactions, but
the \winf theories are entirely local.
We discuss the so-called classical contraction procedure
by which we derive the Lagrangian of \winf gauge theory
from that of the corresponding \Winf gauge theory.
In order to discuss the relationship between quantum \Winf
and quantum \winf gauge theory
we solve $d=1$ gauge theory models of a Higgs field exactly by using
the collective field method.
Based on this we conclude that the \Winf gauge theory can be regarded 
as the large $N$ 
limit of the corresponding $SU(N)$ gauge theory once an appropriate
coupling constant renormalization is made,
while the \winf gauge theory cannot be.

\vfill
\noindent{$^*$e-mail address: kavalov@scisun.sci.ccny.cuny.edu;
\ \ \ \ \ \ \ \ \ sakita@scisun.sci.ccny.cuny.edu}
\eject
%%%%%%%%%%%%%%%%%%%%%%%%%%%%%%%%%%%%%%%%%%%%%%%%%%%%%%%%%%%%%%
\noindent
{\bf 1. Introduction and Summary}

\Winf algebra and its so-called classically
contracted \winf 
algebra appeared recently in various problems in 
physics, in particular in the study of $c=1$ string 
theory [1] and quantum Hall system [2, 3]. 
The gauge theory based on these algebras also
appeared in these studies [4, 5].
In fact the same algebras and the gauge theories
based on them had been proposed previously
as the theories [6, 7, 8] relevant to the large $N$ limit of
$SU(N)$ gauge theories [9].
But it seems to us that not enough studies have been done 
to differentiate the formal and dynamic 
aspect of \Winf and \winf theories.
In view of the recent developments 
we study in this paper this subject as systematically as possible by
using the 
technique developed in the study 
of quantum Hall system [2, 5].

The \Winf algebra is a commutator 
algebra of Hermitian operators
of one harmonic oscillator [10]. It is an infinite-dimensional
Lie algebra. If we choose a set of linearly 
independent real function of
$z$ and $\zbar$ as the parameters of 
\Winf group, the structure constants
of the algebra are expressed in terms of Moyal bracket [11].
Replacing the
Moyal bracket by a Poisson bracket we
define the \winf algebra. It is an algebra
of area-preserving diffeomorphism. 
As an introduction  
we discuss this issue in section 2 together with
the so-called classical contraction
procedure by which the \Winf 
algebra is transformed to the \winf algebra.
 
\Winf gauge 
theory is a gauge field theory of \Winf as an
internal symmetry algebra.
The \Winf gauge potential is a space-time dependent
Hermitian operator of harmonic oscillator. In the coherent state
representation it is a function of $z,\ \zbar$, which
we call the color space coordinates, and $x^\m\ \ (\m =
1, 2,\dots d)$, the space time coordinates.
Thus, we can express
the \Winf gauge theories in terms of $d+2$ dimensional 
local fields. 
The interactions of the 
fields are necessarily non-local in the color space in \Winf 
theories, but they are local in \winf theories.
In section 3 we
define \Winf theories as $d+2$ dimensional field theories 
with non-local interactions and
\winf gauge theories as $d+2$ dimensional
local field theories.
Since the \Winf algebra is closely related to the \winf algebra,
the gauge theories
based on these algebras may also be closely related.      
In order to see the relationship at classical Lagrangian
level, we introduce the $l\to 0$ contraction procedure
by which we derive the \winf gauge theories from the corresponding
\Winf gauge theories.
The procedure consists of the introduction of
a length scale $l$ in the color space, an
appropriate scale transformation of the fields, and the $l\to 0 $ limit.
In this section we also introduce
matter fields analogous to the quark fields and the Higgs fields. 

The \Winf algebra can be considered as the $N\to\infty$ limit of the
$SU(N)$ algebra [10]. Therefore, the \Winf gauge theory or its variation
\winf gauge theory [7, 12] might be used for the large $N$ gauge
theory. 
Since the \winf gauge theory is a local theory
and much easier to be handled, it is important to determine
whether this theory can serve for the large $N$ gauge theory or not. 
For this purpose we 
solve $d=1$ gauge theory of Higgs field exactly in section 4,
which reveals also the quantum mechanical 
relationship between the \Winf
theory and the \winf theory. 
In $d=1$ there exists only the time 
component of the gauge potential and the pure gauge theory is
trivial but it constrains
the states of Higgs field to be gauge
invariant. The \Winf theory becomes essentially
$N=\infty$ limit of $d=1$ gauged matrix model. 
On the other hand the \winf model becomes an infinite
number of non-interacting quantum mechanical systems.
We use the collective 
field method [13] to solve these theories.
The spectrum of these two theories are in 
general different and coincides
with the $N = \inf$ limit of $SU(N)$ theory only for \Winf model.
Based on this result we conclude that \Winf gauge theory can
but \winf gauge theory cannot
serve for the purpose of large $N$ gauge theory. 
%%%%%%%%%%%%%%%%%%%%%%%%%%%%%%%%%%%%%%%%%%%%%%%%%%%%%%%%%%%%%%%%%%

\bigskip
\noindent
{\bf 2. \Winf and \winf Algebra.}  

We define the 
\Winf algebra as a commutator algebra of all 
Hermitian operators 
$\x(\hat{a},\hat{a}^\dag)$ 
in the Hilbert space of a harmonic oscillator [2].
A convenient parametrization for these operators 
is achieved by using a 
real function
$\x (z,\zbar)$ as 
$$
\x (\hat{a},\hat{a}^\dag )
=\ddag\x (z,\zbar ){\bigm | }_{{z=\hat{a}\ }\atop{\zbar =
\hat{a}^\dag}}
\ddag =\int\zintm\ve z \ke \x
 ( z,\zbar )\br z\ve ,
\eqno(2.1)
$$
where $\hat{a}^\dag$ and $\hat a$ are standard 
creation and annihilation operators and   
$\ddag\ \ \ddag$ stands for the anti-normal-order symbol,
i.e., all the creation operators 
stand to the right of the annihilation operators.
The last expression (2.1) is in the coherent 
state representation (see A1).  
Obviously the product of two $\x$'s is not 
anti-normally ordered 
and we bring it to the anti-normal-order form 
by using the commutation relation
$[\hat{a},\hat{a}^\dag]=1$. 
We obtain (see A.1)
$$
\x_1 (\hat{a},\hat{a}^\dag )\x_2 (\hat{a},\hat{a}^\dag )
=\ddag {\sum_{n=1} ^{\infty}}{{(-)^n}\over{n!}}
{\partial_{\zbar} ^{n}}\xi_1 (z,\zbar ) 
{\partial_{z} ^n}\x_2 (z,\zbar ) 
{\bigm | }_{{z=\hat{a}\ }\atop{\zbar =
\hat{a}^\dag}}\ddag ,
\eqno(2.2)
$$
from which the following commutation relation follows
$$
[\x_1 (\hat{a},\hat{a}^\dag ),\x_2 (\hat{a},\hat{a}^\dag ) ]=i
\{\!\!\{ \x_1 ,\x_2 \}\!\!\} (\hat{a},\hat{a}^\dag ) ,
\eqno(2.3)
$$
where $\{\!\!\{ \x_1 ,\x_2 \}\!\!\}$ is a 
Moyal bracket [11] defined by  
$$
\{\!\!\{\xi_1 ,\x_2 \}\!\!\}(z,\zbar)\equiv  
i{\sum_{n=1} ^{\infty}}{{(-)^n}\over{n!}}\left(
{\partial_{z} ^{n}}\xi_1 (z,\zbar ) 
{\partial_{\bar{z}} ^n}\x_2 (z,\zbar )   -
{\partial_{\bar{z}} ^{n}}\xi_1 (z,\zbar )
{\partial_{z} ^n}\x_2 (z,\zbar )\right) .
\eqno(2.4)
$$
The commutation relation (2.3) is that of
the \Winf Lie algebra in the fundamental
representation. The \Winf 
is an infinite-dimensional Lie group with 
parameters being a set of linearly independent
real functions $\x (z,\zbar )$. 
The generators of \Winf are the linear functionals of 
$\x(z,\zbar)$. Thus we write for arbitrary representation:    
$$
[\r [\x_1 ] ,\r [\x_2 ] ]=i\r [
\{\!\!\{ \x_1 ,\x_2 \}\!\!\} ], 
\eqno(2.5)
$$
where $\r$ is the generator of \Winf group.

The Lie algebra 
of \winf , the area-preserving diffeomorphisms,
is defined by the commutation relation
$$
[\r [\x_1 ] ,\r [\x_2 ] ]=i\r [
\{ \x_1 ,\x_2 \} ] ,
\eqno(2.6)
$$
where $\{\ \ ,\ \ \}$ is the Poisson bracket symbol.

It is well known [2] that one can 
obtain the \winf algebra  
from the \Winf by a contraction. 
To explain it 
let us introduce a length scale $l$ in the $z ,\zbar$ space,
which we call the color space,
and set 
$$
z={1\over{{\sqrt 2}l}}(\s_x+i \s_y), \ \ \ \ 
\ \ \ \ \zbar={1\over{{\sqrt 2}l}}(\s_x-i \s_y) . 
\eqno(2.7)
$$
The Poisson bracket is the leading surviving term of the Moyal bracket
in the 
$l\to 0$ limit. To be more specific we set 
$\x (z,\zbar )=l^{-2}\x (\vec\s )$ and obtain 
$$\eqalign{
\lim_{l\to 0} l^{2}
\{\!\!\{\xi_1 ,\x_2 \}\!\!\}(z,\zbar )
&=  \del_{\s_x}\x_1(\vec \s)\del_{\s_y}\x_2(\vec\s) 
- \del_{\s_y}\x_1(\vec\s)\del_{\s_x}\x_2(\vec\s)\cr
&\equiv \e^{ij}\del_i\x_1 (\vec\s )\del_j\x_2 (\vec\s )\cr
&\equiv
\{\x_1,\x_2\}(\vec\s) . \cr} 
\eqno(2.8)
$$
In this paper we call this procedure as the $l \to 0$ contraction. 

%%%%%%%%%%%%%%%%%%%%%%%%%%%%%%%%%%%%%%%%%%%%%%%%%%%%%%%%%%%%%%%%%%%%
\bigskip 
\noindent {\bf 3. \Winf and \winf
Gauge Invariant Lagrangians}

The \Winf gauge 
theory is a gauge field theory of \Winf as an
internal symmetry algebra. 

Let us discuss first the pure Yang-Mills theory.
We introduce a  
gauge potential $\hat{\ba}_\m$  
which is a Hermitian operator in the
harmonic oscillator Hilbert space as well as a function of 
space time:     
$$
\hat{\ba}_\m (x)\equiv\ba_\m (x,\hat{a},\hat{a}^\dag )
=\int\zintm\ve z \ke \ba_\m
 (x, z,\zbar )\br z\ve .
\eqno(3.1)
$$
The action is given by
$$
S_{YM} =-{1\over 4g^2}\int d^d x\  
\tr\big( \hat{\bF}_{\m\n}\hat{\bF}^{\m\n}\big) ,
\eqno(3.2)
$$
where $\hat{\bF}_{\m\n}$ is the field strength defined by
$$
\hat{\bF}_{\m\n} =\del_\m \hat\ba _\n -\del_\n \hat\ba_\m 
-i [\hat\ba _\m ,\hat\ba _\n ] .
\eqno(3.3)
$$

We then rewrite these by using the coherent state representation:
$$
S_{YM} =-{1\over 4g^2}\int\int d^d x d^2 z 
{\sum_{n=0} ^{\infty}}{{(-)^n}\over{n!}}
{\partial_{z} ^{n}}\bF_{\m\n} (x, z,\zbar )
{\partial_{\bar{z}} ^{n}}
 \bF^{\m\n} (x, z,\zbar ) ,
\eqno(3.4)
$$
where
$$
\bF_{\m\n} (x, z,\zbar )=\del_\m \ba_\n 
(x, z,\zbar )-\del_\n \ba_\m (x, z,\zbar )
+\{\!\!\{\ba_\m ,\ba_\n\}\!\!\}(x, z,\zbar ) .
\eqno(3.5)
$$
This action is invariant under the
\Winf gauge transformation:   
$$
\d \hat\ba_\m (x) = \del_\m \hat\x (x) 
+i[\hat\x (x) ,\hat\ba_\m (x)],
\ \ \ \ \ \ \ \
\d\ba_\m (x, z,\zbar )=\del_\m\x (x, z,\zbar ) 
-\{\!\!\{\x ,\ba_\m \}\!\!\}(x, z,\zbar ) .
\eqno(3.6)
$$

In the coherent sate representation the gauge fields are formally
$d+2$ dimensional local fields, $d$ for space time and $2$ for 
color space. However, the interactions are 
non-local since the action
involves derivatives of infinite order.

In the action (3.4) we no longer have damping factor
${\rm e} ^{-|z|^2}$ because it is a trace expression. 
Therefore we have to restrict the field configurations so that
we can integrate by parts in the color space.
In this paper we define
our \Winf gauge theories as $d+2$ dimensional field theories
such that all the fields and their derivatives vanish 
at $z =\infty$.
%%% %%%%%%%%%%%%%%%%%%%%%%%%

Next, let us introduce a fermion field, 
which is a fundamental representation 
of \Winf , namely a field which transforms as
a bra or ket vector in
the Hilbert space of harmonic oscillator:
$$
\ve \y (x)\ke =\int \ve z \ke \zintm\br z\ve \y (x)\ke\equiv\int
\ve z \ke \  \zintm \y (x, \zbar ) .
\eqno(3.7)
$$
We write the action as 
$$
\eqalign{S_F = &\int d^n x \br \y (x) \ve \
\g^\m\big(i\del_\m - \hat{\ba}_\m (x)\big)\ve \y (x) \ke
\cr 
= &\int\int d^n x \zintm \bar{\y} 
(x, z)\g^\m\big(i\del_\m -\ba_\m (x, z,\zbar )\big)\y (x, \zbar ),
}\eqno(3.8)
$$
which is invariant under the \Winf gauge transformation (3.6) and
$$
\d \ve \y(x) \ke =  - i \hat\x (x) \ve \y(x) \ke ,
\ \ \ \ \ \ \ \ \ \ \ \ \ \
\d\y (x, \zbar ) =-i\ddag\x (\del_{\zbar} ,\zbar )\ddag\y (x, \zbar ) ,
\eqno(3.9)
$$
where $\ddag\ \ \ \ddag$ indicates that the derivatives 
are placed on the left of $\zbar$.
%%%%%%%%%%%%%%%%%%%%%%%

As a last example of \Winf gauge theory
let us consider next a 
scalar field which is 
in an adjoint representation. 
In this paper we call this field simply a Higgs field.
$$
\hat M (x)\equiv M(x,\hat a,\hat a ^\dag) =\int\zintm\ve z \ke M
 (x,z,\zbar)\br z\ve .
\eqno(3.10)
$$
The action is given by:  
$$
\eqalign{S_H =&\int d^d x \tr\left[ {1\over 2}
\big( \del_\m \hat{M} (x) 
- [\hat{\ba}_\m ,\hat{M}](x)\big)
\big( \del^\m \hat{M} (x) 
-[\hat{\ba}^\m ,\hat{M}](x)\big)-v(\hat M)\right]\cr 
=& 
\int d^d x \left[ \int d^2 z 
{1\over 2}{\sum_{m=0} ^{\infty}}{{(-)^m}\over{m!}}
{\partial_{z} ^{m}}\big( \del_\m M (x,z, \zbar) 
-\{\!\!\{\ba_\m ,\M\}\!\!\}(x,z,\zbar)\big)\times\right. \cr 
& \ \ \ \ \ \ \ \ \ \ \ \ \ \ \ \ \ \ \ \ \ \left. 
\times {\partial_{\bar{z}} ^{m}} 
\big( \del^\m M (x,z, \zbar)  
-\{\!\!\{\ba^\m ,\M\}\!\!\}(x,z,\zbar)\big) \right]
-\tr  v(\hat M),\cr
&v(\hat M) = \sum_n g_n {\hat M}^n, \ \ \ \ \ \ \ \ \ 
dim(g_n) = d\left({n \over{2}} - 1\right) - n. 
\cr}\eqno(3.11)
$$ 
Here again we require that the fields and their 
$z,\zbar$ derivatives should fall off to zero 
at $z=\inf$. We can check that this action is invariant under the
\Winf gauge transformation (3.6) and 
$$
\d M (x,z,\zbar)= \{\!\!\{\x,M \}\!\!\}(x,z,\zbar),
\ \ \ \ \ \ \ \ 
\d \hat M (x)=-i[\hat\x (x) ,\hat M (x)] .
\eqno(3.12)
$$
Notice that here again the interactions are 
non-local in the color space.

The quantization of the theory is done by the standard canonical
quantization. Although the interactions are non-local in the color
space, they are local in the ordinary space. 
Accordingly there is no problem
for the quantization. We shall see it explicitly in an
example in the next section.
%%%%%%%%%%%%%%%%%%%%%%%%%%%%%%%%%%%%%

Let us discuss next the contraction procedure
which will
allow us to obtain the \winf gauge invariant actions from \Winf
ones. 
For the fields in the adjoint representation 
such as $\ba_\m$ and $M$ this procedure is straightforward.
As we mentioned earlier, 
by introducing two real coordinates $\s_x$ and $\s_y$  
as in (2.6) 
and by taking the $l \to 0$ limit we reduce Moyal bracket 
to Poisson bracket. If we   
simultaneously rescale the fields and the
coupling constants as 
$$
\eqalign{
&\ba_\m(x,z,\zbar)=l^{-2} \ba_\m (x,\vec\s) \cr 
&M(x,z,\zbar) = (2\p )^{1\over 2}l M(x,\vec\s) \cr 
&g^2 =  \tilde{g}^2 l^{-6}, \ \ \ \ 
g_n = \tilde{g}_n {(2\p )^{1\over 2} l}^{2-n} ,
}\eqno(3.13)
$$
we obtain from (3.4) and (3.11) the following  
\winf 
gauge invariant $d+2$ dimensional local field theory:   
$$
\eqalign{
&S_{YM} = -{1\over{4\tilde{g}^2}}\int d^d x d^2 \vec{\s} 
F_{\m\n}(x,\vec\s)F^{\m\n}(x,\vec\s), \cr
&\ \ \ \ \ \ \ \ \ 
F_{\m\n}(x,\vec\s)=\del_\m\ba_\n(x,\vec\s)
-\del_\n\ba_\m(x,\vec\s)  
+\e^{ij}\del_i\ba_\m(x,\vec\s) \del_j\ba_\n(x,\vec\s);\cr
&S_H =
\int d^d x d^2 \vec\s \left[ {1\over 2}
\big(\del_\m M(x,\vec\s) 
- \e^{ij}\del_i\ba_\m(x,\vec\s)\del_j M(x,\vec\s) \big)
\right.\times\cr 
&\ \ \ \ \ \ \ \ \ \ \ \ \ \ \ \ \ \ \ \ \ 
\left. \times \big(\del^\m M(x,\vec\s) 
- \e^{ij}\del_i\ba^\m(x,\vec\s) \del_j M(x,\vec\s) \big)
- \tilde{v} (M)\right] ,\cr
&\ \ \ \ \ \ \ \ \ 
\tilde{v} (M) = \sum_n \tilde{g}_n M^n(x,\vec\s) .
}\eqno(3.14)
$$
Here again we require 
that the fields vanish at $\vec\s = \inf$.

Setting $\x(x,z,\zbar)=l^{-2}\x(x,\vec\s)$ we obtain 
the \winf gauge transformation: 
$$
\eqalign{
\d&\ba^\m(x,\vec\s) =\del ^\m\x(x,\vec\s)  
-\e^{ij}\del_i\x(x,\vec\s) \del_j\ba^\m(x,\vec\s) \cr
\d& M(x,\vec\s) = \e^{ij}\del_i\x(x,\vec\s) \del_j M(x,\vec\s) .\cr}
\eqno(3.15)
$$
Even though the transformations (3.15) are obtained from 
(3.6) and (3.12) by the $l \to 0$ limit, 
we can independently check that
the actions (3.14) is really invariant by the 
\winf gauge transformation (3.15). We remark
that the second equation
of (3.15) can be written as 
$$\d M(x,\vec\s) = M(x,\vec\s +\d\vec\s (x,\vec\s ))
-M(x,\vec\s ),\ \ \ \ \ \ 
\d\s^i (x,\vec\s ) =-\e^{ij}\del_j\x(x,\vec\s),
\eqno(3.16)
$$
which is a local
area-preserving coordinate transformation.
 
As we mentioned earlier the damping factor $e^{-|z|^2}$ 
cancels out in Lagrangians for the fields in the 
adjoint representation such as $\ba_\m$ and $M$,   
due to the property of the trace in coherent state 
representation. But it remains there for the fields 
in fundamental representation, such as Fermi field 
(see (3.8)).  Therefore, a naive 
$l \to 0$ limit leads to the trivial  
result ($S_{F} \equiv 0$).  
This may imply difficulty in introducing a Fermi
field of fundamental representation in \winf theory.
 
We should mention that the YM lagrangian (3.14) 
had already been written down
in the literature [7, 8]. 

%%%%%%%%%%%%%%%%%%%%%%%%%%%%%%%%%%%%%%%%%%%%%%%%%%%%%%%%%%%%% 
\bigskip
\noindent
{\bf 4. One Dimensional Higgs Model: One Dimensional \Winf and \winf
Matrix model}   

In the previous section we presented a general method
for constructing \Winf gauge
theory and then we obtained the \winf gauge theories
by using the $l\to 0$ contraction procedure from \Winf theories.

Several questions arise.
Since \Winf group can be considered as an 
$N=\inf$ limit of $SU(N)$ group, can one
use the \Winf gauge theory for the large $N$ limit
of $SU(N)$ gauge theory,
especially for the large $N$ QCD [9]?
In the large $N$ QCD one takes the $N\to\infty$ limit
keeping $g^2 N$ finite. Since in \Winf theories $N$ is already at
infinity, how can one implement the large $N$ QCD condition ? 
A simple Feynman diagramatic calculation of \winf theory shows that
the coupling constants are multiplicatively renormalized to 
absorb the infinite volume of color space. In a similar way
in \Winf guage theory the question arises whether or not
one can implement the QCD condition as a multiplicative
renormalization of coupling constants ?
As shown in the previous section it is possible to obtain
\winf theory from \Winf theory by contraction.
This shows the relationship between these theories
as classical theories. How about in quantum theory ?
Is there any physical region where
one can use the \winf gauge theory for large
$N$ QCD ? 
In this section we address these questions by solving  
the simplest  
one-dimensional case exactly.

Since in $d=1$ there exists only time 
component of gauge field and
the pure gauge field model becomes trivial, we
consider the gauge invariant Higgs model.
This model may be thought of as a gauged 
one dimensional \Winf (or \winf) matrix model [14].
We solve it by using the collective field method [13].
Since we can carry out the discussions entirely in parallel for 
both \Winf and \winf models,  
we present the corresponding expressions
simultaneously and
put label $a$ for \Winf model and label $b$ for \winf model.

One dimensional \Winf and \winf matrix model Lagrangians
are given by (compare with (3.11) and (3.14)
respectively): 
$$
\eqalign{
{L}=&\tr\left[ {1\over 2}(\del_t \hat{M}-[\hat{\ba_0},\hat
{M}])^2 - v(\hat{M})\right]\cr
{}=&{1\over 2}\int  d^2 z 
\big( \del_t M (t, z, \zbar ) 
-\{\!\!\{\ba_0 ,\M\}\!\!\}(t, z, \zbar )\big)
{\rm e}^{\del_z\del_{\zbar}}
\big( \del_t M (t, z, \zbar ) -\{\!\!\{\ba_0 ,\M\}\!\!\}
(t, z, \zbar )\big)-\tr v(\hat{M}),\cr}
\eqno(4.1a)
$$
$$
{ L}=\int d\vec\s \left[ {1\over 2}\big( \del_t M(t, \vec\s)  
-\e^{ij}\del_i\ba_0(t, \vec\s)\del_j M(t, \vec\s)\big)^2 -v(M) \right] .
\eqno (4.1b)
$$
The canonical quantization leads to the
following Hamiltonians:   
$$
H=\int P (z,\zbar )\del_t M (z,\zbar ;t) d^2 z -L
=  \int d^2 z {1\over 2}\sum^\inf_{n=0} {1\over {n!}}
\del_z ^n P (z, \zbar )\del_{\zbar}^n P(z,\zbar )
+\tr v(\hat{M}) ,
\eqno (4.2a)
$$
$$
H=\int P (\vec x  )\del_t M (\vec x  ;t) d^2 x -L
=  \int d\vec\s \left( {1\over 2} P (\vec\s  )^2
+v (M (\vec\s ))\right) ,
\eqno (4.2b)
$$
where $P$'s are the canonical momentum operators 
conjugated to $M$'s, with the following
commutation relations:  
$$
[\hat{M}(z,\zbar ),\hat{P}(z',\zbar')]=i\d^{(2)}(z-z') ,
\eqno (4.3a)
$$
$$
[\hat{M}(\vec \s ) , \hat{P}(\vec \s ') ]=i\d (\vec\s -\vec\s ' ) .
\eqno (4.3b)
$$
The \Winf (and \winf) gauge invariance 
of the actions (4.1) leads to the following
constraints, which we impose on the state vector 
$|\Y\ke$: 
$$
\hat{\P}_\x |\Y\ke \equiv 
\int d^2 z \{\!\!\{\x,\hat M \}\!\!\}\hat P(z,\zbar)|\Y\ke=0 ,
\eqno (4.4a)
$$
$$
\hat{\P}_\x |\Y\ke \equiv 
\int d\vec\s \e^{ij}\del_i\x (\vec \s)\del_j \hat{M}(\vec \s)\hat
P(\vec \s) |\Y\ke =0 ,
\eqno (4.4b)
$$
which simply state that the wave function (or the state 
vector) is gauge invariant and
depends only on gauge invariant singlet
variables. Therefore in order to solve the problem we
choose the following \Winf (and \winf) invariant
collective field as dynamical variables:   
$$
\f(x) = \tr \d (x - M(\hat{a}, \hat{a}^\dag)) ,
\eqno(4.5a)  
$$
$$
\f (x) =\int d\vec\s \d \big( x - M (\vec\s  )\big) .
\eqno (4.5b)
$$
We then change variables
from $P  (z,\zbar )$, $M(z,\zbar)$ to $\p (x)$, $\f(x)$,
where $\p(x)$ is a canonical momentum 
conjugate to $\f (x)$:
$$
[ \p (x), \f(x)]=-i\d (x-x') +{\rm const.} .
\eqno (4.6)
$$
The standard procedure (see A2) to do this is to compute 
$\Omega (x, x')$ and $\omega (x)$.
We describe this calculation in Appendix A3. 
The definitions and the results are  
$$
\Omega(x, x') \equiv - \int d^2 z 
\sum^\inf_{n=0} {1 \over {n!}}   
[\del_{\zbar}^n P(z, \zbar), \f(x)] 
[\del_z^n P(z, \bar{z}), \f(x')]
=\del_x\del_{x'}[\d (x -x' )\f (x)] ,
\eqno(4.7a) 
$$
$$
\Omega (x,x') \equiv -\int d\vec\s 
[P(\vec \s ),\f (x)][P(\vec \s ),\f (x')]
=\del_x\del_{x'}\big[\d(x-x') \f(x)\big],
\eqno (4.7b)
$$
and 
$$
\w(x) \equiv \int d^2 z  \sum^\inf_{n=0}
{1 \over {n!}}   [\del_{\zbar}^n P(z, \zbar), [ \del_z^nP(z, \zbar ),
\f(x)]]=2\del_x[\f (x) G (x;\f )] ,
\eqno(4.8a)
$$
$$
\omega(x)\equiv \int d\vec \s [P(\vec \s ),[P(\vec \s ),\f (x)]]
=-\k^2 {\del_{x}
^{2}}\f( x ) ,
\eqno (4.8b)
$$
where $\k^2 =\d^2 (0)$, and 
$G(x;\f ) =P\int{{\f (x')}\over {x-x'}}$.   
Notice that we obtained the same expression
for $\Omega$ for both theories but quite 
different expressions for $\omega$.

In the collective field theory the hermiticity requirement of the 
Hamiltonian leads to the
following equation for the Jacobian $J$ of change 
of variables:  
$$
\omega (x)+2\int dx' \Omega (x,x') C(x') =0,
\ \ \ C(x) = {1\over 2 }{\d\over {\d\f (x)}}J ,
\eqno (4.9)
$$
Using (4.7) and (4.8) and assuming $\del_x \f(-\infty) 
=\f (-\infty ) \del_x C (-\infty ) =0 $ we obtain
$$
\del_x C(x) = G(x;\f ) ,
\eqno (4.10a)
$$
$$
\del_x C(x) = -{1\over 2}\k^2 {{\del_x\f (x)}\over{\f (x)}}
=-{1\over 2}\k^2 \del_x  \ln\f(x) .
\eqno (4.10b)
$$

Since the kinetic energy part of the hermitian 
Hamiltonian in the collective 
field theory is given by
$$
K={1\over 2}
\int\int dx dx' \big[\p (x)\Omega (x,x')\p(x')
+C(x)\Omega (x,x')C(x')\big],
\eqno (4.11)
$$
using (4.7) and (4.10) we obtain the following Hamiltonians:
$$
H = \int dx \left( {1 \over 2} {(\del_x \pi (x))}^2 \f(x) 
+ {\pi^2 \over 6} \f(x)^3 
+ v(x) \f(x) \right ) - e 
\left(\int dx\f (x) - N\right) 
\eqno(4.12a)
$$
$$
H= 
\int dx\left( {1\over 2}  \del_x \p (x) \f (x)\del_x \p (x)+
{1\over 8}\k^4 {{ (\del_x\f(x))^2}\over{\f (x)}}
+ \tilde v (x) \f(x) \right) - e 
\left(\int dx\f (x) - L^2\right) 
\eqno (4.12b)
$$
where $e$ is a Lagrange multiplier to insure
$$
\int dx\, \f (x) =\tr 1 \equiv N\ \ \ \ \ 
\ \ \ \ ({\rm for}\ W_\infty ),
\eqno(4.13a)
$$
$$
\int dx\, \f (x)  =\int d\vec\s \equiv L^2\ \ \ \ 
\ ({\rm for}\ \w_\infty)
\eqno (4.13b)
$$
which follows from the definition of collective field (4.5).
Notice that eventually we have to take $N\to\infty,\ \k\to\infty,
\ L\to\infty$.
For this purpose we first make
the following scale transformations:
$$
x\to N^{1/2} x ,\ \ \ \ 
\f (x)\to  N^{1/2}\f (x), \ \ \ \p (x)\to N^{-1}\p (x),\ \ \ \ e\to N e
\eqno (4.14a)
$$ 
$$
x\to \k x ,\ \ \ \ 
\f (x)\to L^2 \k\f (x),
 \ \ \ \p (x)\to L^{-2}\p (x),\ \ \ \ e\to\k^2 e
\eqno (4.14b)
$$ 
which preserve the canonical commutation relation (4.6).
We obtain
$$
H = \int dx \left[ {1\over {2N ^2}} {(\del_x \pi (x))}^2 \f(x) 
+ N^2\left({\pi^2 \over 6} \f(x)^3 
+ u(x) \f(x)-e\f(x)\right )\right] + N^2 e 
\eqno(4.15a)
$$
$$
H= 
\int dx\left[ {1\over {2N ^2}}  {(\del_x \pi (x))}^2 \f(x)+
N^2\left( {1 \over 8} {{ (\del_x\f(x))^2}\over{\f (x)}}
+ \tilde u (x) \f(x)  - e 
\f (x)\right)\right] + N^2 e 
\eqno (4.15b)
$$
where for \winf we set
$
N \equiv L \k 
$
and
$$ 
u(x)=\sum_n N^{\left({n\over2}-1\right)} g_n x^n 
\equiv \sum_n \a_n x^n, 
\eqno(4.16a)
$$
$$
\tilde u (x) =\sum_n \k^{n-2} \tilde g_n x^n
=\sum_n (\k l)^{n-2} g_n x^n 
\equiv \sum_n \tilde \a_n x^n. 
\eqno (4.16b)
$$
We know from the previous study [13, 15] that 
the $1/N$ expansion of Hamiltonians (4.15) 
is a standard semi-classical expansion and in the $N\to\infty$ limit
the excitation spectrum
is finite provided that $u(x)$ is finite and given by
$$
H={1\over 2}\sum_{n=0}^{\infty}\big(p_n ^2 
+\omega_n ^2 q_n ^2\big),
\ \ \ \ \ \ \ \ \ \ [q_n , p_m ]=i\d_{nm},
\eqno(4.17) 
$$
where
$$
\omega_n = n \p / T ,\ \ \ T=
\int_{\tilde{x}_1}^{\tilde{x}_2} 
{{dx'}\over{\sqrt{2(\tilde{e}_0 -u(x'))}}},
\ \ \ \ 
{1\over \p}\int_{\tilde{x}_1}^{\tilde{x}_2} 
{dx'}{\sqrt{2(\tilde{e}_0 -u(x'))}}=1 ,
\eqno(4.18a)
$$
$$
\omega_n = E_n - E_0 , \ \ \ \ \ \ \ \ \ \ \ \
\left( -{1\over 2}\del_x ^2 + \tilde{u}
( x )\right)\chi_n ( x ) =E_n \chi_n (x)
\eqno(4.18b)
$$
The derivation of these results will be given in
detail in Appendix A4 and A5.

The result for
\Winf model exactly coincides with the $N \to \inf$ 
limit of the matix model discussed in [13, 15]. Of cource 
it is expected already from the equations 
(4.7a) and (4.8a) since 
these equations coincide with the ones obtained in [13]. 
The infinite volume of color space (i.e. $N$) is absorbed
by the multiplicative renormalization into the coupling constants
(see (4.16)).

The \Winf effective Hamiltonian (4.15a) is 
actually an element of the \winf spectrum generating
algebra [1]. However this \winf symmetry is not directly related
to the original \Winf gauge symmetry of the Lagrangian,
since the dynamical \winf transformations are realized
non-trivially in the physical Hilbert space
while the original \Winf gauge symmetry acts trivially in the
physical Hilbert space.

As we see in (4.18) the energy spectrum of \Winf model is that of
bosons with equally spaced frequencies irrespective
of its interactions. For the \winf model
the equally spaced spectrum occurs only 
when $\tilde{\a}_n =0 $ for $n\ne 2$,
namely the free \winf theory.
It appears that in quantum theory the 
$l\to\infty$ contraction of \Winf theory is a free
\winf theory. Therefore, it is not possible to
learn \Winf theory by studying \winf theory.

%%%%%%%%%%%%%%%%%%%%%%%%%%%%%%%%%%%%%%%%%%%%%%%%%%%%%%%%%%%%%%
\bigskip
\noindent
{\bf Acknowledgements}

One of us (An.K.) would like to acknowledge
useful discussions with V.P. Nair and G. Alexanian, 
and grateful to the High Energy group for constant
interest in this work and financial support. 
The other (B. S.) would like to thank Y. Nagaoka and K. Shizuya
for their hospitality extended to him at Yukawa Institute
for Theoretical Physics of Kyoto University,
where part of the work was done in the fall
of 1995 under the sponsorship of JSPS Fellowship.
This work was supported by the National Science Foundation, 
grant number PHY - 9420615.     

%%%%%%%%%%%%%%%%%%%%%%%%%%%%%%%%%%%%%%%%%%%%%%%%%%%%%%
\bigskip
\noindent
{\bf Appendix A1. Coherent state representation} 

We list the definitions  and basic 
properties of the coherent state representation 
which we use throughout the paper: 
$$
\eqalign{ 
& \ve z \ke = e^{\hat{a}^\dag z} \ve 0 \ke, \ \ \ \ \ \ 
\br z \ve = \br 0 \ve e^{\hat{a} \zbar}, \ \ \ \ \ \ 
\br z' \ve z \ke = e^{\zbar' z} \cr
& \hat{a} \ve z \ke = z \ve z \ke, \ \ \ \ \ \ 
\br z \ve \hat{a}^\dag = \br z \ve \zbar,  \ \ \ \ \ \ 
\int \zintm \ve z \ke \br z \ve = 1, \ \ \ \ \ \ \ 
d^2 z \equiv {d Re z \  d Im z \over \p} \cr 
}\eqno(A1.1) 
$$

For a given real function $\x (z,\zbar )=\sum_{m,n} \x_{mn}z^n\zbar ^m$
we obtain
$$
\x (\hat a , \hat a ^\dag ) 
=\sum_{m,n} \x_{mn}\hat a ^n  (\hat a ^\dag )^m
=\int\zintm
\sum_{m,n} \x_{mn}\hat a ^n \ve z \ke\br z \ve  (\hat a ^\dag )^m
=\int\zintm\ve z \ke \x
 ( z,\zbar )\br z\ve
\eqno(A1.2)
$$

A proof of (2.2) goes as follows.
$$
\eqalignno{
\x_1 (\hat{a},\hat{a}^\dag )\x_2 (\hat{a},\hat{a}^\dag )
&=\int \zintm \int \zintmp\ve z \ke 
\x_1 (z,\zbar ) \e^{\zbar z'}\x_2 (z',\zbar ' )
\br z '\ve \cr
&=\int \zintm \int \zintmp\ve z \ke \x_1 (z,\zbar ) 
\x_2 (-\overleftarrow{\del}_{\zbar }
,\zbar ' )\e^{\zbar z'}
\br z '\ve\cr
&=\int d^2 z \ve z \ke \x_1 (z,\zbar ) 
\x_2 (z-\overleftarrow{\del}_{\zbar} 
,\zbar ) \e^{-|z|^2}\br z \ve\cr
&=\int \zintm  \ve z \ke{\sum_{n=1} ^{\infty}}{{(-)^n}\over{n!}}
{\partial_{\zbar} ^{n}}\xi_1 (z,\zbar ) 
{\partial_{z} ^n}\x_2 (z,\zbar ) 
\br z\ve
\cr
&=\ddag {\sum_{n=1} ^{\infty}}{{(-)^n}\over{n!}}
{\partial_{\zbar} ^{n}}\xi_1 (z,\zbar ) 
{\partial_{z} ^n}\x_2 (z,\zbar ) 
{\bigm | }_{{z=\hat{a}\ }\atop{\zbar =
\hat{a}^\dag}}\ddag . &(2.2)\cr
}$$
%%%%%%%%%%%%%%%%%%%%%%%%%%%%%%%%%%%%%%%

\noindent
{\bf A2. Change of variables in quantum mechanics  } [13]

We start with a standard form for the Hamiltonian and
Schr\"odinger  equation which is given by
$${
\hat H \psi ~=~ \biggl( ~-~ {1 \over 2} \sum_{a=1}^N ~
{ \partial^2 \over \partial q^{a\, 2} } ~+~
V ( {\bf q} ) \biggr)   \psi ( q ) ~=~ E \psi ( q )
}\eqno (A2.1)
$$ 
We consider a  transformation given by
$$
q^a \longrightarrow ~ Q^a = f^a ( q ),\ \ \ \ q^a ~=~ F^a ( Q )
\eqno(A2.2)$$ 
We use the chain rule of differentiation to convert
the derivatives with respect to $q$'s
into derivatives with respect to $Q$'s.
$${
 -{1 \over 2} \sum_a { \partial^2 \over \partial q^{a\, 2} } 
 \psi ( q )
 ~=~ {1 \over 2} \biggl( ~-~ i \sum_{ a b }
{ \partial^2 f^b \over \partial q^{a\, 2} }
{\partial \over  i\partial Q^b } 
 ~+~  \sum_{ abc } ~ { \partial f^b \over \partial q^a } 
 { \partial f^c \over \partial q^a } 
 {\partial \over i\partial Q^b }
{\partial \over i\partial Q^c } \biggr) \psi ( F ( Q ) ) 
}\eqno(A2.3)$$ 
We define
$$
\omega^a ( Q ) \equiv - \sum_b  
{ \partial^2 Q^a \over \partial q^{b\, 2} } 
=- \sum_b { \partial^2 f^a \over \partial q^{b\, 2} },
\ \ \ \ 
\Omega^{ ab } ( Q ) \equiv~ \sum_c 
 { \partial Q^a \over \partial q^c }
 { \partial Q^b \over \partial q^c } =  \sum_c 
{ \partial f^a \over \partial q^c } 
 { \partial f^b \over \partial q^c } .
\eqno(A2.4)$$ 
Then we obtain
$${
\hat H \psi ~=~ \biggl[ {i \over 2} 
\bigl( i \sum_a \omega^a ( Q ) P_a
 ~+~ \sum_{ a b } \Omega^{ a b } ( Q ) P_a P_b \bigr)
 ~+~ \tilde V ( Q ) \biggr] \psi ( F ( Q ) ) ,
}\eqno(A2.5)$$ 
where we set
${1 \over i} {\partial \over \partial Q^a } ~=~ P_a ,\ \ \ \ \ 
\tilde V ( Q ) ~\equiv~ V ( F ( Q ) ).$

The Hamiltonian after the change of variables appears to be
non-Hermitian if we take the naive Hermitian conjugate:
${
{P_a}^{\dag}= P_a  , 
{Q^a}^{\dag} =Q^a 
}$.
This is because $ H $ is Hermitian in the original $ q $-space.
But after the change of variables,  the $ Q $-space should be defined by
multiplying the wave function by
the square root of the Jacobian 
in order to satisfy the naive Hermitian conjugation prescription:
$${
\int d q \, \psi_1^* ( q ) \psi_2^{} ( q ) ~=~
\int  J ( Q ) \, d Q \, \psi_1^* ( F ( Q ) ) \,
\psi_2^{} ( F ( Q ) ) 
 =~ \int d Q \, \Psi_1^* ( Q ) \, \Psi_2^{} ( Q )  , 
}\eqno(A2.6)$$ 
where
$
\Psi ( Q ) =J^{1/2} ( Q ) \psi ( F ( Q ) ).$ 
The Hamiltonian in $Q$-space is then obtained by a
similarity transformation
$${
H_{\rm  eff } ~=~ J^{1/2}  H  J^{ - {1/2} },
}\eqno(A2.7)$$ 
which should be Hermitian.

In practice the Jacobian is difficult to calculate while
$ \omega $ and $ \Omega ~$ defined by (A2.4) are
relatively easy to compute.  So, it would be nice if $ H_{ \rm eff } $
is expressed in terms of $ \omega $ and $ \Omega  .$   Notice first
$
J^{\dag} ( Q ) = J ( Q^{\dag} ) = J ( Q ),$ 
accordingly
$
J^{1/2} P_a J^{ - {1/2} } =P_a
+ i C_a ( Q ),
$
where
$
C_a ( Q ) ={1 \over 2} {\partial \over \partial Q^a }
\ln J ( Q )$
and
$
  C_a^{\dag}= C_a$.
We obtain
$$
 H_{\rm eff } = {1 \over 2} \bigl[ ~i \sum_a \omega^a ( Q )
( P_a ~+~ i C_a )
+ \sum_{ ab } \Omega^{ ab }
( P_a ~+~ i C_a )  ( P_b ~+~ i C_b ) \bigr]
+ \tilde V ( Q ) .
\eqno(A2.8)
$$
Since $ H_{ \rm eff } $ should be Hermitian,
$
H_{\rm eff } -H_{\rm eff }^{\dag} =
i \sum_a \bigl\{ ( \omega^a ~+~ 2 \sum_b \Omega^{ ab }
C_b ~+~ \sum_b \Omega_{~,b }^{ ab } ) , P_a \bigr\}_+ = 0 ,
$ 
and by taking a commutator bracket with $ Q_a $ we obtain
$$
\omega^a ~+~ 2 \sum_b \Omega^{ab} C_b
 ~+~ \sum_b \Omega_{ ~,b }^{ab}    ~=~ 0 .
\eqno(A2.9)$$
This is the equation that determines $ C_a $.   $ H_{\rm eff } $
is then computed
as 
$$
H_{ \rm eff  } ~=~ {1 \over 2} \sum_{ ab } \bigl[ P_a
\Omega^{ ab } P_b ~+~ C_a \Omega^{ab}
C_b ~+~( C_a \Omega^{ab} )_{, b } \bigr] ~+~ \tilde V
\eqno(A2.10)$$ 
(A2.9) and (A2.10) are the main results.

%%%%%%%%%%%%%%%%%%%%%%%%%%%%%%%%%%%%%%%%%%%% 
\bigskip
\noindent
{\bf A3. Calculation of $\Omega $ and $\omega$}

In order to calculate $\Omega(x,x')$ 
and $\omega(x)$ for \Winf theory one needs the 
following equations:   
$$
\eqalign{
&[P(z,\zbar),
e^{- i k M(\hat{a}, \hat{a}^\dag)}]
=- e^{-|z|^2} k \int_0^1 d \tau
e^{- i k \tau M(\hat{a}, \hat{a}^\dag)} \ve z\ke 
\br z\ve e^{- i k (1-\tau) M(\hat{a}, \hat{a}^\dag)} ,\cr
&[P(z, \zbar), \f(k)] = - k e^{- |z|^2}
\br z\ve e^{- i k M(\hat{a}, \hat{a}^\dag)} \ve z \ke ,\cr   
&[P(z, \zbar), [P(w, \bar{w}), \f(k)]] = k^2 
e^{-|z|^2 - |w|^2} \int_0^1 d \tau
\br w\ve e^{- i k \tau M(\hat{a}, \hat{a}^\dag)} \ve z\ke 
\br z\ve e^{- i k (1-\tau) M(\hat{a}, \hat{a}^\dag)} \ve w \ke .
}\eqno(A3.1)
$$
The proof is straightforward
once we realize 
the following identity:
$$
\e ^{-|z|^2}\sum_n {1\over{n!}} 
\big( (\del_z -\zbar )^n\ve z \ke \big) 
\big(  (\del_{\zbar }- z )^n \br z \ve \big) = 1 . 
\eqno(A3.2)
$$
The (A3.2) expresses the completeness property
of the generators of \Winf fundamental representation. 
It can be considered as a generalization
of the completeness property of $SU(N)$ 
generators to the case of
$N=\inf$ . 
One can prove this identity by  
multiplying both sides by 
arbitrary ket vector $\ve z'\ke$. Then the right hand side is 
$\ve z' \ke$
and the left hand side is equal to: 
$$
\eqalign{&\e ^{-|z|^2}\sum_n{1\over{n!}}(\hat{a}^\dag 
-\zbar )^n\ve z \ke 
(\del_{\zbar} - z )^n \e^{\zbar z'}\cr
&= \sum_n{1\over{n!}}\big((\hat{a}^\dag -\zbar )
(z' -z)\big)^n\ve z \ke \e^{-|z|^2 +\zbar z'}\cr
&=\e^{(\hat{a}^\dag -\zbar )(z' -z)}\ve z \ke \e^{-|z|^2 +\zbar z'}\cr
&=\e^{\hat{a}^\dag z'}\e^{-\hat{a}^\dag z}\ve z\ke\cr
&=\ve z'\ke \ \ \ \ \ \ \ \ \ \ \ {\rm (QED)}}
\eqno(A3.3)
$$
 
%%%%%%%%%%%%%%%%%%%%%%%%%%%%%%%%%%%%%%%

\bigskip
\noindent 
{\bf A4. Solution of \Winf matrix model} 

Although we can solve this model starting from (4.15a)
in a straightforward fashon [15]
as we shall discuss it in A5 for \winf model, in this Appendix we 
solve it in a slightly different way which 
illuminates the dyanmical group
structure of the theory.

We start with the Hamiltonian (4.12a) with the constraint (4.13a):
$$
H = \int dx \left( {1 \over 2} {(\del_x \pi (x))}^2 \f(x) 
+ {\pi^2 \over 6} \f(x)^3 
+ v(x) \f(x) \right ) ,\ \ \ \ \ 
\int dx\f (x) = \tr 1 =N .
\eqno(A4.1)
$$
Using
$
y_{\pm}(x) =\pm\p\f(x) -\del_x\p (x)$
one [16] writes the Hamiltonian and the constraint as
$$
H ={1\over{2\p}}\int dx \int_{y_- (x)}^{y_+ (x)} 
dy\left( {1\over 2} y^2 +v(x)\right) ,
\ \ \ \ \ \ \ \ {1\over{2\p}}\int dx \int_{y_- (x)}^{y_+ (x)} dy=N ,
\eqno(A4.2)
$$
where $y_{\pm} (x) $ 's satisfy the commutation relation
$$
[y_{\pm} (x), y_{\pm} (x') ]=\mp 2\p i\d ' (x-x').
\eqno(A4.3)
$$

We diagonalize this Hamiltonian by 
a canonical transformation. In the integral of (A4.2)
we change variables from $ y, x $ to
$e,\x$ such that
$
e ={1\over 2} y^2 +v(x) .
$
The action integral is the generator of transformation
$
S(x,e)=\int ^x ydy =\pm\int^x \sqrt{2(e - v(x'))} dx' ,
$
accordingly
$
\x ={{\del S}\over{\del e}}=\pm \int^x 
{{dx'}\over {\sqrt{2(e - v(x'))}}},
$
where $\pm$ is for positive and negative $\x$ respectively.
The boundary of the phase space is transformed from $y_{\pm}(x)$
to $e(\x )$
$$
e(\x ) ={1\over 2} y_{\pm}^2 (x) +v(x) \eqno(A4.4)
$$
and
the Hamiltonian and the constraint are given by
$$
H={1\over{2\p}}\oint d\x \int^{e(\x )} ede ,
\ \ \ \ \ 
{1\over{2\p}}\oint d\x \int^{e(\x )} de 
=N .\eqno(A4.5)
$$

We expand $e(\x )$ around the minimum configuration $e_0$ of $H$:
$
e(\x ) = e_0 +\d e(\x ),
$
where
$e_0$ is given by a solution of
$$
{1\over \p}\int_{x_1}^{x_2}\sqrt{2 (e_0 -v(x))}dx =N ,
\eqno(A4.6)
$$
where $x_1$ and $x_2$ are the turnning points.

Since we see from (A4.6) that $e_0$ is of order $N$, we may assume
$e_0 \gg \d e(\x )$ and we may approximate $\x$ on the boundary by
$
\x (x) = \pm \int_{x_1}^x {{dx'}\over{\sqrt{2(e_0 -v(x'))}}}
$
and the half period by
$$
T=\int_{x_1}^{x_2} {{dx'}\over{\sqrt{2(e_0 -v(x'))}}}
=\int_{\tilde{x}_1}^{\tilde{x}_2} 
{{dx'}\over{\sqrt{2(\tilde{e}_0 -u(x'))}}},
\eqno(A4.7)
$$
where $\tilde{e}_0=N^{-1}e_0 ,\ \  
\tilde{x}_1= N^{-{1\over 2}} x_1 $ and 
$u(x)= N^{-1}v (N^{1\over 2} x) $ (see (4.16a)).

Since the commutation rules of $e(\x )$'s are given by
$$\eqalign{
[e(\x ), e(\x ' ) ]&=\left[ {1\over 2} y_{\pm} ^2 (x ) +v (x) ,
{1\over 2} y_{\pm} ^2 (x' ) +v (x') \right]\cr
&=\mp 2\p i y_{\pm} (x) y_{\pm } (x' ) \d ' (x -x' )\cr
&\approx -2\p i \d '(\x -\x '),\cr}
$$
if we define $\theta$ and $r(\theta )$ by
$
\theta= \omega \x ,\ \ r (\theta) = \omega ^{-1}\d e(\x ) , \ \ \ \omega
={\p\over T},
$
we obtain
$$
[r (\theta ) , r (\theta ') ] = -2\p i \d ' (\theta -\theta ' )
\eqno(A4.8)
$$
The normal mode expansion of $r(\theta )$ is given by
$
r (\theta ) =\sqrt{ 2\over\omega }\sum_{n>0}
( \sin ( n\theta ) p_n + n\omega\cos ( n\theta ) q_n ),
$
and leads to the following Hamiltonian:  
$$
\eqalign{  
&H= E_0 + H_{\rm coll}, \cr 
&H_{\rm coll } = \oint{d\x \over {2\p}}
\int_0^{\d e (\x )} ede =\omega \oint {d\theta\over{2\p}}
{1\over 2} (r(\theta ))^2 = 
{1\over 2}\sum_n (p_n ^2 +\omega _n ^2 q_n ^2 ) ) ,
\cr &\omega_n =n\omega = n {\p\over T}.
}\eqno(A4.9)
$$ 
From (A4.6) and (A4.7) it is obvious that 
$T^{-1}$ is finite in the large $N$ limit
provided $u(x)$ is finite. In the double 
scaling limit one of the turning
points goes to infinity so that $T\to \infty$ and we obtain the 
continuous spectrum (chiral Boson).

%%%%%%%%%%%%%%%%%% dynamical \winf algebra %%%%%%%%%%%%%%%%%%%%

The reason why we could solve the \Winf model 
by a canonical transformation
is that there exists a dynamical \winf algebra 
in the physical Hilbert space
and the Hamiltonian is a generator of the algebra. We simply
quote a result of [2], namely $\r [\x]$'s defined by
$$
\r [\x ]=\oint {{d\theta}\over{2\p}}\int ^{r (\theta)} 
dr \x (r,\theta ),
\eqno(A4.10)
$$
satisfy the \winf commutation relation (2.6).

%%%%%%%%%%%%%%%%%%%%%%%%%%%%%%%%%%%%%%%%%%%%%%%%%%%%%%%%%%
\bigskip
\noindent
{\bf A5. \winf Matrix Model}

We first obtain the field configuration 
at which the potential
energy of (4.15b) is minimum. The equations are
$
-{1\over 4}\del\left({{\del\f }
\over{\f}}\right) 
-{1\over 8}\left({{\del\f }\over{\f}}\right)^2  
+\tilde{u}(x) =e
$
and
$
\int d x 
\f (x) = 1 .
$
For the variable  
$
\vf(x) =\sqrt{\f(x)}
$
the first equation is the Scr\"odinger  equation
$
\left( -{1\over 2}\del^2 + \tilde{u}
( x )\right)\vf (x) =e \vf (x)
$
and the second equation is the normalization of the wave function.
Therefore we consider an orthonormal set of eigenfunction 
$\h_n$ with eigenvalue $E_n$.
We set 
$
\f ( x ) = \chi^{2} _{0}( x ) + {1\over{\sqrt N}} 
\j ( x ),\ \ \ \ \p (x ) ={\sqrt N}\z ( x )
$
and expand the Hamiltonian. We obtain
$$
H ={1\over 2}\int d x \left[ \chi^2 _0 ( x )
\big(\del \z ( x )\big)^2 +{1\over 4}\left(
 {{(\del \j ( x ) )^2}
\over{\chi^2 _0 ( x )}}+(2\del^2 \ln\chi_0 )
{{\j^2 ( x )}\over {\chi^2 _0 (x )}}\right) \right]
\eqno(A5.1)
$$
$\j ( x )$
and $\z (x)$ satisfy the  
canonical commutation relation:
$$
[\del \z ( x ) , \j (x ' ) ] = -i \d '( x -  x' )
\eqno(A5.2)
$$
and the constraint
$
\int d x \j ( x ) =0.
$
The normal mode expansion is now straightforward:
$$
\j ( x ) =\chi_0 ( x )\sum_{n=1}^{\infty}
\sqrt{2\omega_n}\chi_n ( x ) \, q_n ,\ \ \ \ 
\z ( x )=\chi^{-1} _0 ( x )\sum_{n=1}^{\infty}
{1\over{\sqrt{2\omega_n}}}\chi_n (x)\, p_n .
\eqno(A5.3)
$$
and
$$
H={1\over 2}\sum_{n=0}^{\infty}\big(p_n ^2 +\omega_n ^2 q_n ^2\big),
\ \ \ \ \ \ 
[q_n , p_m ]=i\d_{n,m}\ ,\ \ \ \ \ \ \omega_n =E_n - E_0 .
\eqno(A5.4)
$$
%%%%%%%%%%%%%%%%%%%%%%%%%%%%%%%%%%%%%%%%%%%%%%%%%%%%
\bigskip
\noindent
{\bf References}
\item {[1]}
J. Avan and A. Jevicki, {\it Phys. Lett.} {\bf B266} (1991) 35;
{\it Phys. Lett.} {\bf B272} (1991) 17; 
\item{} A. Gerasimov, A. Marshakov,
A. Mironov, A. Morozov and A. Orlov, 
{\it Nucl. Phys.} {\bf B357} (1991) 565;
\item {} D. Minic, J. Polchinsky and Z. Yang, 
{\it Nucl. Phys.} {\bf B362} (1991) 125; 
\item {} G. Moore and N. Seiberg, 
{\it Int. J. Mod. Phys.} {\bf A8} (1992) 2601; 
\item {} I. Klebanov and A.M. Polyakov,
{\it  Mod. Phys. Lett.} {\bf A6} (1991) 3273;
\item {}E. Witten, {\it Nucl. Phys.} {\bf B373} (1992) 187;
\item {}S.R. Das, A. Dhar, G. Mandal and S.R. Wadia, 
Int. J. Mod. Phys. {\bf A7} (1992) 5165;
\item {[2]}S. Iso, D. Karabali and B. Sakita , {\it Phys. Lett.} 
{\bf B296} (1992) 143
\item {[3]}  A. Cappelli, C. Trugenberger and G. Zemba,
{\it Nucl. Phys.} {\bf B396} (1993) 465
\item{[4]}  S.R. Das, A. Dhar, G. Mandal and S.R. Wadia, 
{\it Int. J. Mod. Phys.} {\bf A7} (1992) 5165; 
\item {} A. Dhar, G. Mandal and S.R. Wadia, 
{\it Mod. Phys. Lett.} {\bf A7} (1992) 937;
\item {[5]}
B. Sakita , {\it Phys. Lett.} {\bf B315} (1993) 124
\item{} K. Shizuya, {\it Phys. Rev.} {\bf B52} 
(1995) 2747 
\item{[6]} E. G. Floratos and J. Iliopoulos, 
{\it Phys. Lett.} {\bf B201} (1988) 237;
\item {} E. G. Floratos, {\it Phys. Lett.} {\bf B228} (1989) 335
\item {[7]} 
E. G. Floratos, J. Iliopoulos and G. Tiktopoulos, 
{\it Phys. Lett.} {\bf B217} (1989) 285;
\item {[8]} D. B. Fairlie, P. Fletcher and C. K. Zachos, 
{\it Phys. Lett.} {\bf B218} (1989) 203;
\item {} D. B. Fairlie and C. K. Zachos, 
{\it Phys. Lett.} {\bf B224} (1989) 101;
\item {} D. B. Fairlie, P. Fletcher and 
C. K. Zachos, {\it J. Math. Phys.}
{\bf 31} (1990) 1088;
\item {} C. K. Zachos {\it Hamiltonian Flows, 
$SU(\infty ), SO(\infty ), USp(\infty )$,
and Strings} in 
{\it Differential Geometric Methods inTheoretical Physics; 
Physics and Geometry},
NATO ASI Series, L.-L. Chau and W. Nahm (eds.), 
Plenum, New York, p.423, 1990
\item {[9]} G. 't Hooft, {\it Nucl. Phys.} 
{\bf B72} (1974) 461; {\it Nucl. Phys.} 
{\bf B75} (1974) 461
\item {[10]} See [2] [3]. For earlier work see
\item {} J. Hoppe, MIT Ph. D. Thesis (1982); 
\item {} J. Hoppe and P. Schaller, {\it Phys.
Lett.}{\bf B237} (1990) 407;
\item {} C.N. Pope, L.J. Romans and X. Shen
{\it ``A Brief History of $W_{\infty }$''} in 
Strings 90, ed. R. Arnowitt et al
(World Scientific 1991) and references therein.
\item {[11]} J. E. Moyal, 
{\it Proc. Camb. Phyl. Soc.} {\bf 45} (1949) 99
\item {[12]} I. Bars, {\it Phys. Lett.} {\bf B245} (1990) 35 
\item {[13]}
A. Jevicki and B. Sakita, {\it Nucl. Phys.} 
{\bf B165} (1980) 511. See also
\item {}
B. Sakita, {\sl "Quantum Theory of Many-Variable Systems and Fields"},
World Scientific, Singapore (1985)
\item {[14]} S. J. Rankin, {\it Ann. Phys.} {\bf 218} (1992) 14
\item {[15]} M. Mondello and E. Onofri, 
{\it Phys. Lett.} {\bf 98B} (1981) 277   
\item {} R. Jackiw and A. Strominger, 
{\it Phys. Lett.} {\bf 99B} (1981) 133
\item {} J. Shapiro, {\it Nucl. Phys.} {\bf B184} (1981) 218
\item {[16]} J. Polchinski, {\it Nucl. Phys.} {\bf B362} (1991) 125 
\end